\documentclass[%
 reprint,twocolumn,floatfix,
 amsmath,amssymb,
 aps,
prl,
superscriptaddress
]{revtex4-2}

\usepackage{graphicx}
\usepackage{dcolumn}
\usepackage{bm}
\usepackage{color}
\usepackage{comment}
\usepackage[normalem]{ulem}
\usepackage{amsmath}
\usepackage{amssymb}

\newcommand{\figurewidth}{.45\textwidth}
\renewcommand{\epsilon}{\varepsilon}

\newcommand{\tn}[1]{\textnormal{#1}}  

\begin{document}

\title{Discontinuous transition in electrolyte flow through charge-patterned nanochannels}

\author{Tine Curk}
\email{tcurk@jhu.edu}
\affiliation{Department of Materials Science \& Engineering, Johns Hopkins University, Baltimore, Maryland 21218, USA}

\author{Sergi G. Leyva}
\author{Ignacio Pagonabarraga}
\affiliation{Departament de F\'{i}sica de la Mat\`{e}ria Condensada, Universitat de Barcelona, 08028 Spain}
\affiliation{Universitat de Barcelona Institute of Complex Systems (UBICS), Universitat de Barcelona, 08028 Spain} 


\begin{abstract}

 We investigate the flow of an electrolyte through a rigid nanochannel decorated with a surface charge pattern. Employing lattice Boltzmann and dissipative particle dynamics methods, as well as analytical theory, we show that the electro-hydrodynamic coupling leads to two distinct flow regimes. The accompanying discontinuous transition between slow, ionic, and fast, Poiseuille flows is observed at intermediate ion concentrations, channel widths, and electrostatic coupling strengths. These findings indicate routes to design nanochannels containing a typical aqueous electrolyte that exhibit a digital on/off flux response, which could be useful for nanofluidics and ionotronic applications.

\end{abstract}

\maketitle

Ion transport through nanochannels often exhibits non-linear effects such as gating and pressure sensing~\cite{perozo2002,sukharev2004,bavi2016}. These mechanisms are generically present in biological nanochannels.
For example, channels can adapt their shape in response to mechanical stresses and act as emergency safety valves to avoid cellular damage~\cite{haswell2011}, or change  the fluid flows to trigger electrochemical signals~\cite{martinac2002}.
Much effort has been made to mimic the capabilities of such biological mechanisms.
For example, conical nanopores with constant surface charge exhibit gating as a function of the exerted pressure, and have been extensively characterized, both experimentally and theoretically~\cite{siwy2004,harrell2005,cervera2007,jubin2018,lin2019,dalcengio,vanroij23}. 
Such geometrical asymmetries can result in rectification~\cite{siwy2004,harrell2005} and
particle separation due to entropic transport~\cite{malgarettiek}. Both molecular sized pores and nanochannels can give rise to gating and rectification, but the precise response and the physical mechanisms at play may change drastically~\cite{gosh2018,xie2023,zhu2019,jubin2018,lin2019,dalcengio,cervera2007}.

An alternative avenue to obtain non-linear response is by introducing charge heterogeneities. 
Theoretical investigations indicate that a discontinuity in the surface charge causes a disturbance in the flow profile that can extend a distance from the surface an order of magnitude larger than the Debye screening length~$\lambda_{\tn D}$~\cite{khair2008}. 
Indeed, surface charge patterns in micron-sized channels can result in intricate electroosmotic flows~\cite{stroock2000}, and complex flow patterns such as vortex formation that enhance fluid mixing~\cite{datta_2013,lee_2006,ghosh2012}.
Hence, surface charge patterns can qualitatively alter electrokinetic flows, opening up the
possibility to exploit this feature  to control ionic transport in nanochannels.

\begin{figure}
\centering
\includegraphics[width=\figurewidth]{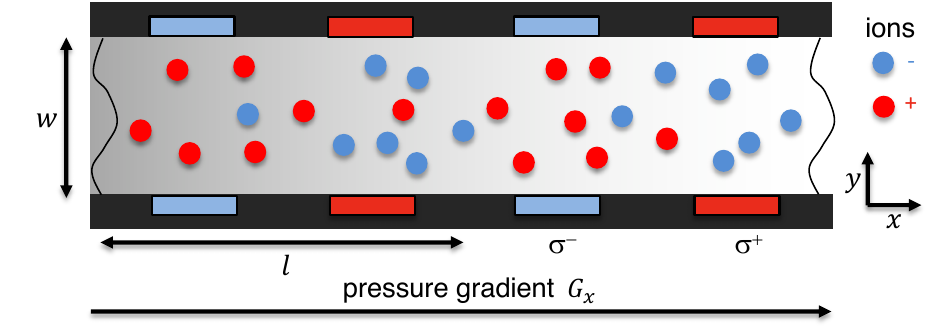}
\caption{Schematic of an electrolyte flow through a charge-patterned nanochannel under a pressure gradient~$G_x$. Channel width~$w$ with a charge pattern of size~$l$ of alternating positive~$\sigma^+$ and negative~$\sigma^-$ charge density. 
}
\label{fig:scheme}
\end{figure}

Here we investigate the flow of an electrolyte through a nanochannel slit of width $w$ at low Reynolds numbers ($\tn{Re}\ll1$). 
In the absence of charge, the flow through a channel with slip length~${\ell}_{s}$ attains the parabolic, Poiseuille velocity profile 
\begin{equation}
v^\tn{P}_x(y) = \frac{G_x}{2\eta} \left(w^2/4 - y^2 + w{\ell}_{\tn s}\right) \;,
\label{eq:Pois}
\end{equation}
where $G_x$ is the pressure gradient in the $x$-coordinate, $\eta$ the dynamic viscosity and the channel walls are positioned at $y=\pm w/2$.
Charging the surface modifies this flow profile due to electrokinetic coupling between the hydrodynamic flow and electrostatic interactions. The electrostatic interaction strength is controlled by the Bjerrum length~$l_\tn{B}=e_0^2/(4\pi\varepsilon_0\varepsilon_\tn{r}k_\tn{B}T)$, with $e_0$ the elementary charge, $\varepsilon_0$ and $\varepsilon_\tn{r}$ the vacuum and relative permittivity, $k_\tn{B}$ the Boltzmann constant and $T$ the absolute temperature, which yields a typical length $l_\tn{B}=0.71\,$nm for an aqueous solution at room temperature. We employ the simplest charge pattern that preserves charge neutrality; an alternating pattern of positive and negative charged sections with pattern size~$l$ and fraction of the surface~$f$ with symmetric surface charge density, $\sigma^+=-\sigma^-=\sigma$ (Fig.~\ref{fig:scheme}). The channel contains an electrolyte solution at density~$\rho$ and monovalent ion concentration~$c_\tn{ion}$. To investigate how the surface charge affects the flow we employ two independent computational methods that combine hydrodynamics with electrostatics, Lattice Boltzmann (LB) with electrokinetics~\cite{capuani2004} and Dissipative-particle dynamics (DPD)~\cite{groot1997} with explicit ions. The analysis is further supported by analytical mean-field theory. 

We initially focus on the parameters corresponding to a channel width $w=5.16\,$nm, containing an aqueous salt solution ($\rho=10^3\,\tn{kg}/\tn{m}^3$, $\eta=10^{-3}\,\tn{Pa}\,\tn{s}$, $T=293\,$K) and consider a typical slip length~${\ell}_s$ for electrolytes on surfaces with different degree of polarity, which is generically small, ${\ell}_s\approx20\,$nm~\cite{joseph2005}. The surface charge pattern length is $l=10w$ with $f=0.5$, which makes the width of each charged strip comparable to the width of the channel (Fig.~\ref{fig:scheme}). The pattern charge density is $\sigma \approx \pm0.5e_0/\tn{nm}^2$, which is typical of, e.g., silica or iron oxide surfaces~\cite{trefalt16}.

\begin{figure}
\centering
\includegraphics[width=\figurewidth]{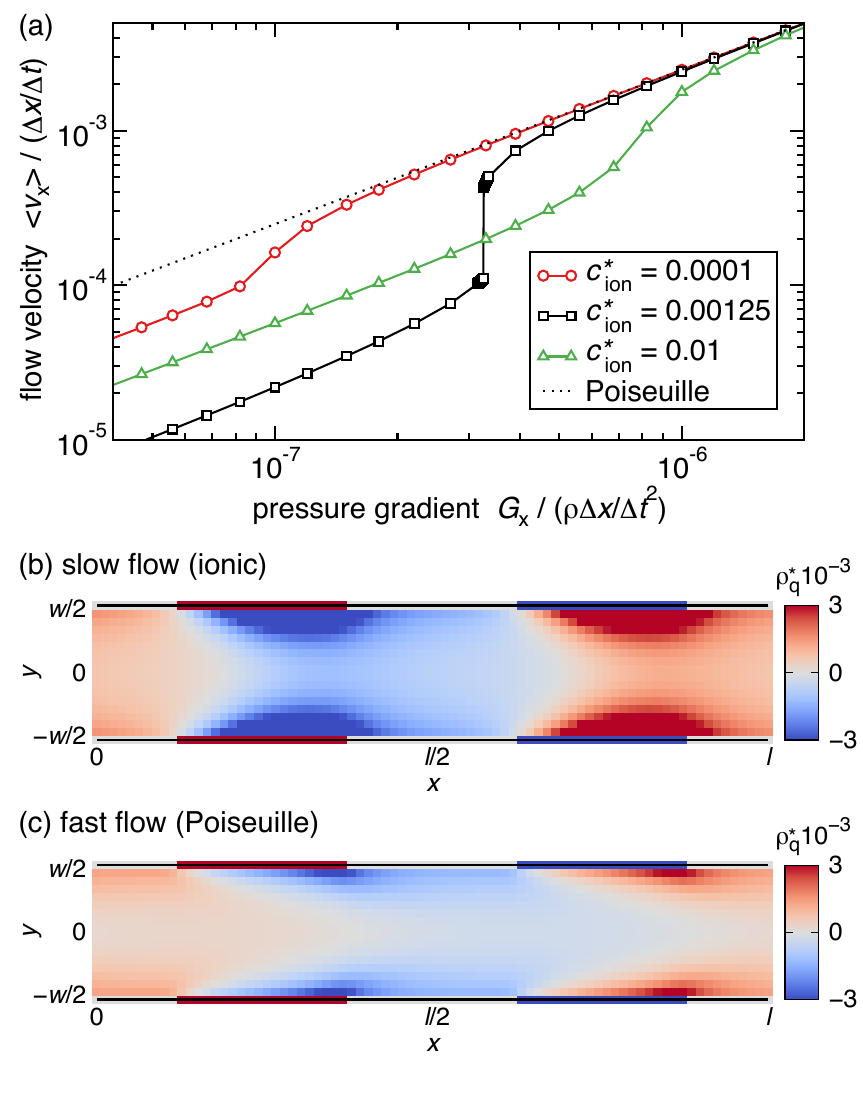}
\caption{Steady-state flow velocity obtained from LB simulations. (a) average velocity at different ion concentrations, $c^*_\tn{ion}=c_\tn{ion} \Delta x^3$. The dashed line corresponds to ideal Poiseuille flow [Eq.~\eqref{eq:vP}]. (b,c) net charge density, $\rho^*_\tn{q}=\rho_\tn{q} \Delta x^3/e_0$, at coexistence conditions for the (b) slow and (c) fast flow profiles at $G=3.25\cdot 10^{-7}\rho\Delta x/\Delta t^2$ and $G=3.26\cdot 10^{-7}\rho\Delta x/\Delta t^2$, respectively, and $c^*_\tn{ion}=0.00125$.}
\label{fig:lb}
\end{figure}

The LB method combined with a convection--diffusion solver for ions~\cite{capuani2004} allows us to analyze the steady-state flow of an electrolyte as a function of pressure gradient~$G_x$ and ion concentration~$c_\tn{ion}$ (Fig.~\ref{fig:lb}). To model small ion diffusion with typical diffusion constant $D\approx 10^{-9}\tn{m}^2/\tn{s}$ we set $D=10^{-3} \nu$, with the kinematic viscosity $\nu=\eta/\rho$. The slip length is introduced using a fractional bounce-back boundary condition at the walls~\cite{wolff2012}. The lattice size is set to $\Delta x=w/16$, which is sufficiently small to avoid finite size effects (see SI), and the reduced viscosity is set to $\eta^*=0.2$, which determines the LB time unit $\Delta t=\Delta x^2 \eta^*/\nu$. 

Surprisingly, we find that at a threshold pressure gradient~$G_\tn{t}$, the flow velocity exhibits a discontinuous transition characterized by nearly an order of magnitude change in the average flow velocity (Fig.~\ref{fig:lb}a). This transition is associated with a discontinuous change in the ion distribution measured by the net charge density~$\rho_\tn{q}$ in the channel.
The slow flow regime shows localized counterion  clouds that reflect the surface charge pattern (Fig.~\ref{fig:lb}b), whereas the charge density is largely uniform in the fast flow regime with only a scant signature of the counterion layer (Fig.~\ref{fig:lb}c). This suggest that at $G_x< G_\tn{t}$ the counterions are localized in a pattern reflecting the surface charge, which results in a high drag on the fluid and thus a distinct slow flow regime. Conversely, at $G_x>G_\tn{t}$, the drag becomes sufficiently large to pull the counter-ion away from the patterned surface charge, leading to ion mixing and associated reduction in local net charge density, which in turn substantially reduces the ion drag and results in a discontinuous transition.  For $G\gg G_\tn{t}$, electrokinetic effects become negligible and the average flow velocity is determined by the Poiseuille flow,
\begin{equation} 
v_\tn{P}=\langle v^\tn{P}_x(y) \rangle = \frac{G_xw^2}{12 \eta} \left(1+\frac{6 {\ell}_s}{w}\right)\;.
\label{eq:vP}
\end{equation}

The discontinuous transition is only observed at intermediate ion concentrations, whereas both higher and lower salt concentrations  result in a non-linear, but continuous flow dependance on $G_x$.  This peculiar behavior is a consequence of many-body electrokinetic effects. At low ion concentrations, $c_\tn{ion}\to 0$, electrostatic interactions become irrelevant and the flow attains the Poiseuille profile. In the opposite limit the electrostatic effects become confined to a narrow boundary layer since $\lambda_{\tn D} \propto c_\tn{ion}^{-1/2}$ and thus the flow again approximately follows a Poiseuille profile. Conversely, at intermediate ion concentrations the electrokinetic effects can qualitatively change the flow, leading to a discontinuous transition that separates the fast and slow flow regimes. The same argument implies the transition only occours at intermediate Bjerrum lengths since $\lambda_{\tn D} \propto l_\tn{B}^{-1/2}$.

Although the LB calculations clearly point to a discontinuous transition in the flow, the method does not include thermal fluctuations and assumes a continuous charge distribution. 
To establish whether the observed transition is affected by thermal fluctuations or unit charge discretization, we turn to DPD, which is an off-lattice method that models the solvent as a fluid of soft particles and allows the introduction of explicit ions.

\begin{figure}
\centering
\includegraphics[width=\figurewidth]{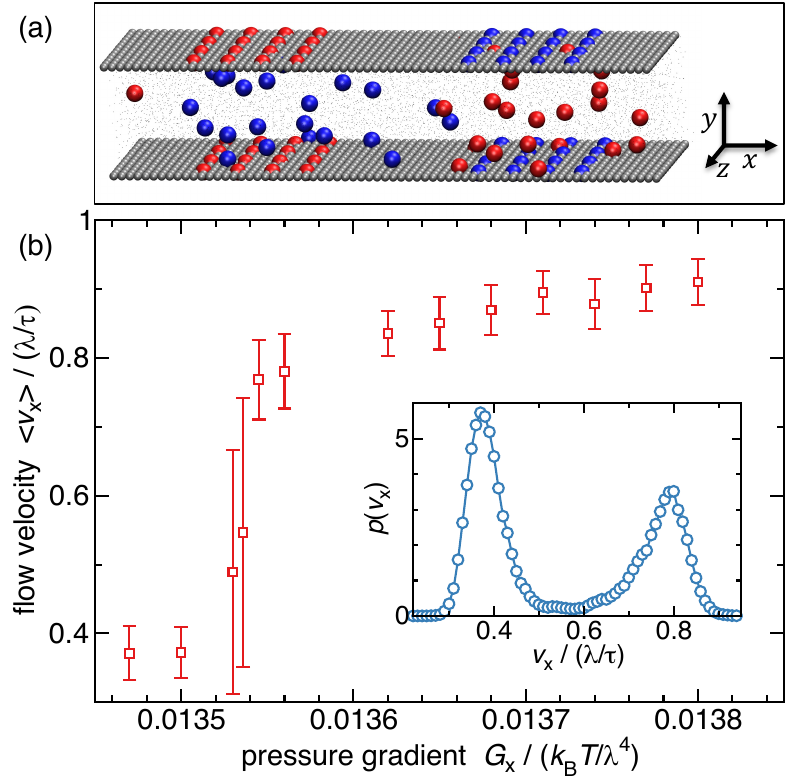}
\caption{Steady state flow from DPD simulations. (a) Nanochannel configuration where cations and anions are shown as red and blue spheres, respectively while channel walls are shown in grey with embedded surface charges (at $G=0.003 k_\tn{B}T/\lambda^4$). DPD particles are represented as small black dots. (b) Average velocity, error bars mark the standard deviation of the velocity distribution. Inset shows the velocity distribution at coexistence ($G=0.004512 k_\tn{B}T/\lambda^4$). System size $L/\lambda=[40,8,8]$ and $c_\tn{ion}=0.01\lambda^{-3}$. }
\label{fig:dpd}
\end{figure}

We use standard DPD parameters corresponding to an aqueous solution~\cite{groot1997, boromand2015} with DPD particle density $\rho_\tn{s}=3/\lambda^2$, at $\lambda=0.645\,$nm and hydrodynamic coupling $\gamma=4.5k_\tn{B}T\tau/\lambda^2$.
We introduce electrolyte ions as charged spheres with diameter $\lambda_\tn{ion}=\lambda$  (Fig.~\ref{fig:dpd}a). The short-range ion--ion and ion--wall repulsion is modeled using the standard WCA potential with strength $\varepsilon=k_{\tn B}T$. The same WCA form is used to describe the smooth channel wall interaction with DPD particles and ions. To separate thermodynamic and hydrodynamic parameters, ions and DPD particles have no conservative pair-interaction and interact only through the DPD thermostat~\cite{curk2023dpd}. 
The ion--DPD hydrodynamic coupling is set to $\gamma_\tn{ion}=5\gamma$, which yields the desired diffusion constant of ions, $D\approx \tn{nm}^2/\tn{ns}$, where the time unit, $\tau=0.077\,$ns, is determined from reduced viscosity~$\eta_\tn{dpd}^*=0.85$~\cite{boromand2015} via $\eta=\eta_{\tn{dpd}}^*k_{\tn B}T \tau \lambda^{-3}$. 
The pressure gradient is introduced as an external body-force on the solvent DPD particles. The partial-slip boundary condition at the walls is implemented by introducing immobilized particles at the wall with surface density $\rho_{\tn w}=\rho_{\tn s} \lambda$ that interact with DPD particles only via the thermostat with coupling~$\gamma_\tn{w}$, which is determined by the desired slip-length~$\ell_\tn{s}$.
Electrostatic interaction are calculated using PPPM Ewald summation (see SI for details). 

\begin{figure}
\centering
\includegraphics[width=\figurewidth]{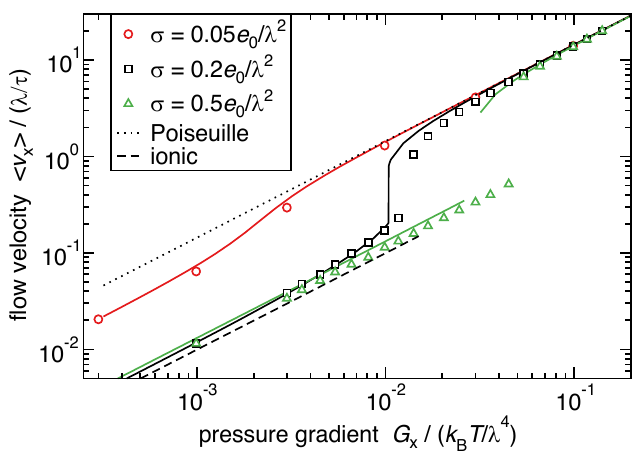}
\caption{Steady state flow at different surface charge densities~$\sigma$. Comparison between DPD (symbols), LB (solid lines), and analytical prediction for the Poiseuille regime (dotted line, Eq.~\eqref{eq:vP}) and ionic regime at $\sigma=0.2e_0/\lambda^2$ (dashed line, Eqs.~\eqref{eq:vD} and~\eqref{eq:Gt}). Parameters: $c_\tn{ion}=0.01\lambda^{-3}$. 
}
\label{fig:compare}
\end{figure}

 Using this DPD model we find that the steady-state flow in a charge-patterned nanochannel exhibits a doubly-peaked velocity distribution (Fig.~\ref{fig:dpd}b) which implies a discontinuous transition between slow and fast flow regimes. Moreover, the transition becomes sharper at higher surface charge densities, in quantitative agreement with LB  (Fig.~\ref{fig:compare}). The agreement is remarkable given that LB does not account for either thermal fluctuations or discrete charges. This indicates that the existence and the location of the discontinuous transition is robust and is not sensitive to the details of the model.

Based on the simulation results, we propose a mean-field  theory that captures the essential features of the electro-hydrodynamic coupling. The flow velocity is determined by both the drag of ions localized in the channel and the viscous drag of the walls. Specifically, the viscous drag force of the two confining walls~$F_\tn{w}$ is determined by the shear rate at the walls,
\begin{equation}
F_\tn{w}= \pm2 A \eta \left(\frac{\partial v_x}{\partial y}\right)_{y=\mp w/2} \;,
\end{equation}
with $A$ the surface area of the wall.
The Stokes drag per ion is
 $f_{\tn i} = -3 \pi \eta \lambda_\tn{ion} v_\tn{i}$,
 where $v_\tn{i}$ is the velocity of the ion relative to the surrounding fluid and $\lambda_\tn{ion}$ is the hydrodynamic diameter of the ions.
At small pressure gradients, ions are confined to the charged regions (Fig.~\ref{fig:dpd}a) thus the relative velocity is, $v_\tn{i}=-v_x(y)$. The total ion drag~$F_\tn{i}$ on the fluid is obtained by integrating over all counter ions in the charged channel  section. Approximating that the counter-ion charge density $\rho_\tn{q}$ does not vary with $x$ within each charged section,
\begin{equation}
F_{\tn{i}} = 3 \pi \eta \lambda_\tn{ion} f A \int_{-w/2}^{w/2} v_x(y) \frac{\rho_\tn{q}(y)}{e_0} \tn{d}y \;,
\label{eq:Fion}
\end{equation}
where $\rho_\tn{q}(y)/e_0$ is the counterion concentration profile. 
For $\lambda_\tn{D}\ge w/2$ the two screening layers from the opposite walls overlap and $\rho_\tn{q}(y)$ is approximately uniform and determined by the surface charge density for $c_\tn{ion}> f\sigma/(w e_0)$ or the overall ion concentration for $c_\tn{ion}< f\sigma/(we_0)$, $\rho_\tn{q}(y) =\min[2\sigma/w ,c_\tn{ion}e_0]$.
Conversely, for  $\lambda_\tn{D}< w/2$ and $c_\tn{ion}> f\sigma/(we_0)$, the surface charge pattern is screened (Fig.~\ref{fig:lb}) and the ion contribution becomes negligible, $F_{\tn{i}}\sim 0$.


We can now analytically determine the ratio of drag forces due to bound counterions and the channel walls. Assuming the profile remains parabolic,
\begin{equation}
R=\frac{F_{\tn{i}}}{F_\tn{w}}= \frac{\pi \lambda_\tn{ion} w}{2} \min[\sigma f, w c_\tn{ion}e_0] \left(1+\frac{6 {\ell}_\tn{s}}{w}\right) \;, 
\label{eq:R}
\end{equation}
which is independent of the pressure gradient. 
The average flow velocity in the channel $\langle v_x \rangle \propto G_x/(F_\tn{w} + F_\tn{ion})$ can be written as,
\begin{equation}
\langle v_x \rangle= \frac{v_\tn{P}}{1+R}\;.
\label{eq:vD}
\end{equation}
 For $R\ll1$ the ion contribution is negligible and the flow attains the Poiseuile profile [Eqs.~\eqref{eq:Pois} and~\eqref{eq:vP}], whereas for $R>1$ the ion drag dominates and we call this regime ``ionic".

The transition between the two flow profiles will occur when the drag force is sufficiently large to pull the ions away from the charge pattern. This force can be estimated analytically by approximating the charge distribution with a point charge~$Q_\tn{s}$ per surface patch depth $w$ (the relevant lengthscale), $Q_\tn{s}=\sigma w f l$ (Fig~\ref{fig:scheme}), and a corresponding point charge~$Q_i$ for the counterions in the center of the channel, while neglecting interactions beyond $w$.  The net counterion charge is determined by either the surface charge, or the ion concentration if ions cannot fully compensate the surface charge, $Q_i=\min[Q_\tn{s}, c_\tn{ion}w^2le_0]$.
The resulting maximum restoring force is $F_{\tn{max}} = k_\tn{B}T \frac{8Q_s Q_i l_\tn{B}}{3\sqrt{3}w^2 e_0^2}$ and the transition between the diffusive and Poiseuille flow regimes occurs at a pressure gradient that can overcome this force, $G_{\tn t}=2F_{\tn{max}}/(w^2l)$, which equals to
\begin{equation}
G_{\tn t} =  \frac{16k_\tn{B}T l_{\tn B}\sigma f l c_\tn{ion}\min[1,f\sigma/(c_\tn{ion}we_0)]}{3^{\frac{3}{2}} we_0} \;.
\label{eq:Gt}
\end{equation}

\begin{figure}
\centering
\includegraphics[width=\figurewidth]{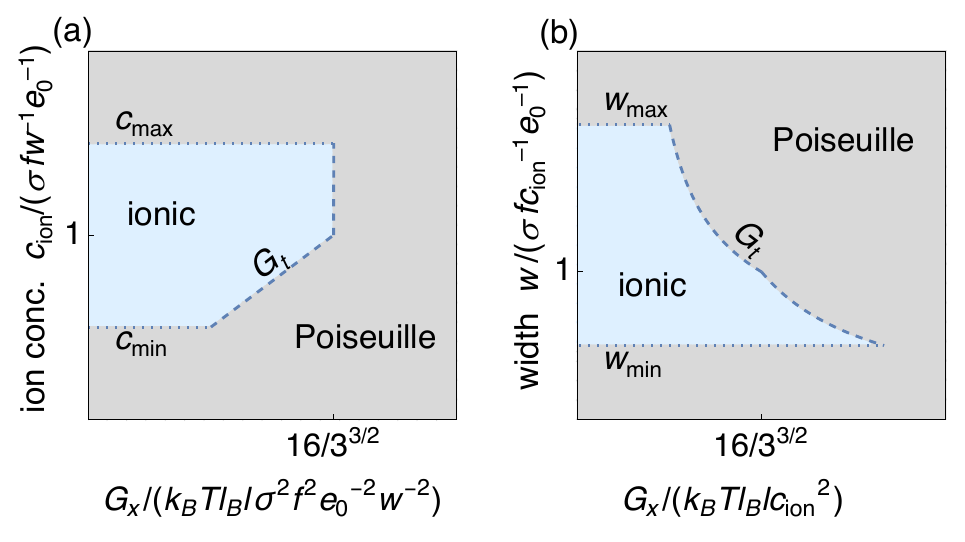}
\caption{Diagrams delineating the Poiseuille and diffusive flow regimes: (a) constant $c_\tn{ion}$, (b) constant  $w$. The upper bounds $c_\tn{max}$ and $w_\tn{max}$ are determined by $w = 2\lambda_\tn{D}$, while the lower bounds $c_\tn{min}$ and $w_\tn{min}$ are determined by $R=1$ [Eq.~\eqref{eq:R}] and shown as dotted lines. The transition at $G_\tn{t}$ (dashed line) is determined by Eq.~\eqref{eq:Gt}.}
\label{fig:pd}
\end{figure}

This theory is able to semi-quantiatively predict both the flow velocity~$\langle v_x\rangle$ and the location of the transition~$G_{\tn t}$ (Fig.~\ref{fig:compare}).
Moreover, the theory predicts general regions of parameter space where different flow regimes are expected to be observed (Fig.~\ref{fig:pd}). The ionic regime is found only at intermediate $c_\tn{ion}$ and $w$, while its extent depends on the surface charge~$\sigma$ and slip length~$\ell_\tn{s}$. 
The larger the slip length, the larger the relative drag of ions [Eq.~\eqref{eq:R}] and thus the larger the jump at the transition [Eq.~\eqref{eq:vD}].

\begin{figure}
\centering
\includegraphics[width=\figurewidth]{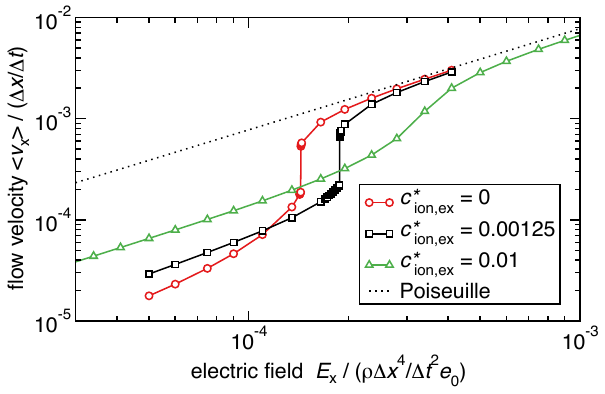}
\caption{Steady-state flow velocity for el. field-driven flow at $\sigma=0.1 e_0/\Delta x^2$. The channel configuration is the same as shown in Fig.~\ref{fig:lb}], but without the negative surface charge. The dashed line corresponds to ideal Poiseuille flow [Eq.~\eqref{eq:vP}] at force density $G_x=E_x \langle \rho_q \rangle$. Theory [Eq.~\eqref{eq:Gt}] predicts the transition at $E_\tn{t}\approx2\cdot10^{-4}\rho \Delta x^4/(\Delta t^2e_0)$.}
\label{fig:elf}
\end{figure}

For non-neutral charge patterns the fluid attains a net charge and the electrosmotic flow can be induced by an external electric field $E_x$ instead of a pressure gradient. For a pattern consisting of only one polarity we again observe a discontinuous transition (Fig.~\ref{fig:elf}). The only notable difference is a lower limit for the counterion concentration $\langle \rho_q \rangle=2f\sigma/w$, at which the transition remains discontinuous even in the absence of extra salt density~$c_\tn{ion,ex}$. 
The net body force, $G_x=E_x \langle \rho_q \rangle$, is determined by the net charge density~$\langle \rho_q \rangle$ and the location of the transition, 
$E_\tn{t} = G_\tn{t}/\langle \rho_q \rangle$, is semi-quantitatively determined by the theory [Eq.~\eqref{eq:Gt}]. Thus, we expect the flow-regime-diagrams (Fig.~\ref{fig:pd}) are qualitatively applicable to flows driven by electric fields. 

The sharp flow transition is reminiscent of the ionic Coulomb blockade effect that occurs at the level of individual ions or electrons at small channel widths $w\lesssim \l_\tn{B}$ and leads to sharp changes in the ionic current depending on the surface charge density. 
However, Coulomb blockade is limited to strong electrostatic coupling and does not predict a discontinuous transition as a function of electrostatic field~\cite{kavokine2019}. 
Therefore, we conclude that the observed phenomenon (Fig.~\ref{fig:elf}) is 
distinct from the Coulomb blockade effect.

In summary, we have investigated the flow of an electrolyte solution through a rigid nanochannel decorated with a surface charge pattern and demonstrated the capability of effective gating for overall electroneutral channels. Simulation results and analytical theory predict two distinct flow regimes, a slow ion-drag dominated flow, and a faster Poiseuille flow,  separated by a discontinuous transition. This transition occurs only at intermediate ion concentrations, channel widths, and electrostatic coupling strengths and appears to be qualitatively different both from the Coulomb blockade effect~\cite{kavokine2019} in nanochannels and the continuous laminar--turbulent transition in pipe flow~\cite{sano2016}.

While mechanosensitive nanochannels are common in biology, their non-linear response is typically coupled to structural changes in the channel such as protein conformational changes~\cite{cox2019}. Our findings imply that such structural changes are not necessary to obtain two distinct (on/off) flow profiles.  
Moreover, the principles that drive the discontinuous flow transition open venues for the design of nanochannel devices, an alternative to those based on conical pores~\cite{jubin2018,vanroij23} and angstrom-scale slits~\cite{lyderic2021}, that could also result in a memristive response. Hence, the possibility to control ionic transport through charge-patterned nanochannels make them potential components in iontronics and the design of brain-inspired neuronal circuits.~\cite{vanroij23}.

\begin{acknowledgments}
T.C. thanks Erik Luijten for enlightening discussions and acknowledges support from Whiting School of Engineering (JHU) through startup funds, and from Advanced Research Computing at Hopkins (rockfish.jhu.edu), which is supported by the National Science Foundation (NSF) grant number OAC 1920103.  I.P. acknowledges support from Ministerio de Ciencia e Innovaci\'on  MICIN/AEI/FEDER for financial support under grant agreement PID2021-126570NB-100 AEI/FEDER-EU, and from Generalitat de Catalunya under Program Icrea Acad\`emia and project 2021SGR-673.
 \end{acknowledgments}


%

\end{document}